# TBI Image/Text (TBI-IT): Comprehensive Text and Image Datasets for Traumatic Brain Injury Research

Jie Li[2], Jiaying Wen[1], Tongxin Yang[2], Fenglin Cai[2], Miao Wei[3], Zhiwei Zhang[3], and Li Jiang[*1]

[1]Department of Neurosurgery, the First Affiliated Hospital of Chongqing Medical University, Chongqing, PR China

[2]Chongqing University of Science and Technology, Chongqing, PR China

[3]Department of Radiology, the First Affiliated Hospital of Chongqing Medical University, Chongqing, PR China

**Abstract**

In this paper, we introduce a new dataset in the medical field of Traumatic Brain Injury (TBI), called TBI-IT, which includes both electronic medical records (EMRs) and head CT images. This dataset is designed to enhance the accuracy of artificial intelligence in the diagnosis and treatment of TBI. This dataset, built upon the foundation of standard text and image data, incorporates specific annotations within the EMRs, extracting key content from the text information, and categorizes the annotation content of imaging data into five types: brain midline, hematoma, left cerebral ventricle, right cerebral ventricle and fracture. TBI-IT aims to be a foundational dataset for feature learning in image segmentation tasks and named entity recognition.

Index Traumatic Brain Injury, Deep learning, Named Entity Recognition, Novel Dataset

## I. INTRODUCTION

Traumatic brain injury (TBI) is one of the most common emergencies in neurosurgery, with the characteristics of high morbidity and high mortality rate, especially for young adults. Meanwhile, the condition not only becomes a burden for the families of those affected but also presents challenges to national healthcare systems[1]. Research on how to improve the efficiency of TBI treatment and curb its harm has significant social benefits and economic value[2]. Therefore, rapidly and accurately identifying the key clinical manifestation and analysising the head CT images are crucial to the diagnosis and treatment.

Despite some models can automatically segment head CT images, challenges remain in terms of precision and contextual understanding[3]-[7]. Meanwhile, there is a scarcity of meticulously annotated head CT and text datasets related to TBI，which is an important challenge for the research and application of TBI intelligent diagnosis and treatment models. TBI-IT is a Chinese database for TBI. And is dedicated to support open research in the fields of head CT image segmentation and EMR extraction for patients with TBI:

1) We provide a dataset of TBI images and EMR named entities..

2) The dataset is comprehensively described in terms of its origin, composition, format, and labels.

3) The CT image and EMR of the same patient in this dataset were matched.

## II. RELATED WORKS

The dataset originates from authentic medical records, in order to protect patient privacy, we have excluded sections that are closely related to patient privacy from the original records. The dataset consists of two distinct annotated datasets: one for head CT images and the other for EMRs. The image data, annotated using 3D-Slicer, is segmented into NIfTI format slice images. The annotation data for the images consists of five parts: brain midline, hematoma, left cerebral ventricle, right cerebral ventricle and fracture. The positional information of all five parts has been comprehensively annotated. The text data includes detailed text information related to the patient's medical history, current condition, physical examination findings, laboratory results, and examination reports, all of which are closely linked to subsequent diagnosis and treatment. These information has been manually annotated using methods based on the BERT model. This is done to facilitate text information processing approaches such as Named Entity Recognition (NER).

This dataset plays a crucial role in interdisciplinary research, particularly in the fields of medicine, computer science, and artificial intelligence. With the limited availability of extensively annotated datasets for TBI head CT images and EMRs, our objective is to create this dataset to drive progress in intelligent healthcare. The dataset enables experimentation with models for image segmentation and text recognition. Moreover, it helps clinical professionals quickly understand patient conditions, ultimately improving diagnostic and treatment effectiveness. In the field of artificial intelligence, this dataset provides a wealth of experimental data for the development of more advanced medical image processing and EMR analysis algorithms. This contributes to achieving more accurate and effective methods for medical diagnosis and treatment [8]–[11].

## III. TBI-IT DATASET

*A. Data Source*

The foundation of any impactful medical research lies in the quality and relevance of the data sources used. Our objective is to utilize and continuously update this dataset for the long term. Therefore, the source of the data is of paramount importance. The EMR and head CT image data in the TBI-IT are sourced from the emergency electronic medical record database of several medical centers. The image data consists of head CT scans completed during patients's visits to the emergency department, while the text data consists of text data in electronic medical records.

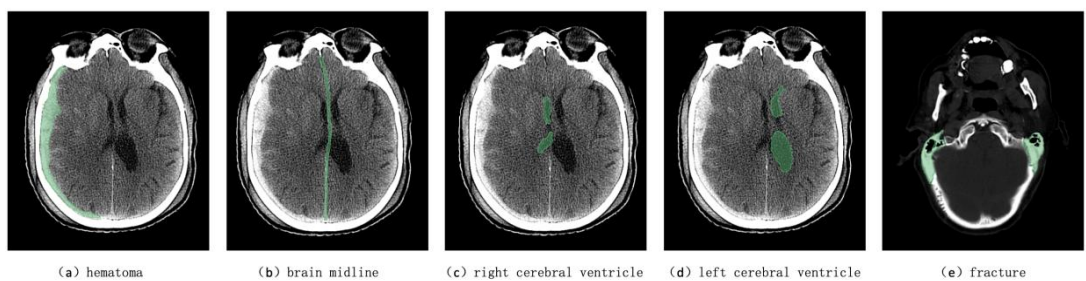

Figure 1 Partial annotation information for the head CT images

*B. Dataset Composition*

This dataset encompasses several hundred thousand CT images and thousands of case texts. Additionally, due to the extensive volume of text information, we have selected some of the information to show, the images are represented as viewed within 3D-Slicer. This dataset is composed of head CT images and EMR texts, both carefully annotated. The annotation information for the head CT images is divided into five categories: brain midline, hematoma, left cerebral ventricle, right cerebral ventricle and fracture, as illustrated in Figure 1. The case text information has been annotated for significant sections using methods based on the BERT model, as depicted in Figure 2. In addition, since the text data is stored as a whole column in the text file, it is necessary to make formatting adjustments in order to achieve a better display effect.

```
2 B-LOC 9 I-LOC 岁 I-LOC 女 I-LOC 性 I-LOC ， O 主 O 诉 O ： O 车 B-LOC 祸 I-LOC 致 I-LOC 头 I-LOC 痛 I-LOC 7 I-LOC 天 I-LOC 。 O 现 O 病 O 史 O ： O 7 B-LOC 天 I-LOC 前 I-LOC ， O 患 O 者 O 驾 B-LOC 驶 I-LOC 车 I-LOC 辆 I-LOC 不 I-LOC 慎 I-LOC 追 I-LOC 尾 I-LOC 致 I-LOC 头 I-LOC 部 I-LOC 外 I-LOC 伤 I-LOC ， O 诉 O 头 B-LOC 痛 I-LOC ， O 无 O 双 O 眼 O 黑 O 朦 O 、 O 意 O 识 O 丧 O 失 O ， O 未 O 做 O 特 O 殊 O 处 O 理 O 。 O 病 B-LOC 程 I-LOC 中 I-LOC 患 I-LOC 者 I-LOC 头 I-LOC 痛 I-LOC 无 I-LOC 改 I-LOC 善 I-LOC ， O 为 O 求 O 进 O 一 O 步 O 外 O 科 O 治 O 疗 O ， O 遂 O 于 O 我 O 院 O 就 O 诊 O 。 O 既 O 往 O 史 O ： O 无 O 。 O 个 O 人 O 史 O ： O 否 O 认 O 疫 O 区 O 、 O 疫 O 情 O 、 O 疫 O 水 O 接 O 触 O 史 O ， O 否 O 认 O 吸 O 烟 O 史 O ， O 否 O 认 O 饮 O 酒 O 史 O ， O 否 O 认 O 放 O 射 O 性 O 物 O 质 O 及 O 化 O 学 O 毒 O 物 O 接 O 触 O 史 O 。 O
```

Figure 2 Partial annotation information for the text data

*C. Data Format*

Due to that medical image data comprises four key components: pixel depth, photometric interpretation, metadata, and pixel data. Therefore, the preferred format for storing medical datasets is typically the NIfTI format. NIfTI images are often three-dimensional, representing sagittal, coronal, and axial planes upon slicing. The advantage of this format lies in its ability to accurately reflect metadata, including directional information, making it highly suitable for neurosurgical-related image data [12]–[14]. Our dataset's head image data consists of several hundred thousand CT images, which are stored in NIfTI format. Each image slice measures 512*512 pixels. The text data, on the other hand, comprises tens of thousands of cases and is stored in TXT format.

## V. CONCLUSIONS

This paper presents a meticulously annotated dataset tailored for advancing research in the relatively underrepresented domain of TBI image and associated case text information extraction. The dataset is applicable for hot-topic issues like image segmentation and natural language processing, providing a wealth of experimental data for the development of medical image processing and case text analysis algorithms. To demonstrate the versatility of the dataset, we provide an overview of its origin, format, and composition, among other relevant details. We believe that this dataset holds the potential for excellent performance with more refined models and advanced algorithms, contributing significantly to the development of intelligent healthcare solutions.The CT image and EMR of the same patient in this dataset were matched. This is expected to enhance the accuracy and efficiency of clinical diagnoses and treatments.

We aimed to update this dataset and expand both the quantity and variety of the dataset regularly. Progress and results were reported in due course.